\documentclass{article}
\usepackage{spconf,graphicx,IEEEtrantools,cite}
\usepackage{amsmath,amsfonts,amsthm,amssymb, bm}
\usepackage{enumitem,kantlipsum}
\usepackage{tikz}
\usetikzlibrary{shapes,arrows,positioning,calc}

\usepackage[linesnumbered,ruled,vlined]{algorithm2e}

\title{A Novel Bayesian Approach for the Two-Dimensional Harmonic Retrieval Problem}
\name{Rohan R. Pote and Bhaskar D. Rao\thanks{This research was supported by ONR Grant No. N00014-18-1-2038
and the UCSD Center for Wireless Communications.}}
\address{Department of Electrical and Computer Engineering\\ University of California San Diego}
\begin{document}
\ninept
\maketitle
\begin{abstract}
Sparse signal recovery algorithms like sparse Bayesian learning work well but the complexity quickly grows when tackling  higher dimensional parametric dictionaries. In this work we propose a novel Bayesian strategy to address the two dimensional harmonic retrieval problem, through \emph{remodeling} and \emph{reparameterization} of the standard data model. This new model allows us to introduce a block sparsity structure in a manner that enables a natural pairing of the parameters in the two dimensions. The numerical simulations demonstrate that the inference algorithm developed (H-MSBL) does not suffer from source identifiability issues and is capable of estimating the harmonic components in challenging scenarios, while maintaining a low computational complexity. 
\end{abstract}
\section{Introduction}\label{sec:intro}
Many signal processing applications involve solving a multidimensional harmonic retrieval (MHR) problem. Examples include MIMO radar \cite{nion10}, MIMO wireless channel sounding \cite{liu05}, and nuclear magnetic resonance spectroscopy \cite{li98}. In addition to resolution problems, a new challenge for methods in 2-D is that
though the pairs are distinct, there could be harmonics
with the same value in one dimension, i.e. $(a,b)$ and $(a,c).$ To address the MHR problem, many techniques have been proposed in the past \cite{hua90, pesavento04, roy89,zoltowski96,haardt98, schmidt86,trees02,gershman10,liu06}, as well as efforts were made to gain a solid theoretical understanding \cite{sorenson17,jiang01}. Subspace based methods for solving the MHR problem include multiple signal classification (MUSIC) \cite{schmidt86,trees02}, estimation of signal parameters via rotational invariance technique (ESPRIT) \cite{roy89,zoltowski96,haardt98}, rank reduction estimator (RARE) \cite{pesavento04}, multi-dimensional folding (MDF) \cite{liu05} and matrix pencil \cite{hua90}. Many of these methods rely on the assumption of uncorrelated sources, and their performance degrades when the assumption does not hold. Also techniques like ESPRIT are applicable to uniform array geometries only. Approaches like \cite{hua93,pesavento04}{\color{red}} extend 1-D techniques to solve higher dimensional estimation problem, such that an additional coupling of parameters in each dimension needs to solved. On the other hand, techniques in \cite{haardt98,liu06} allow a natural pairing of the two sets of parameters.

In this work we approach the two-dimensional harmonic retrieval problem using sparse signal recovery (SSR) tools. SSR methods allow one to reconstruct a spectrally sparse signal using highly incomplete temporal or spatial data \cite{candes06,donoho06}. In particular, in this paper we address the higher computational complexity associated with the sparse Bayesian learning (SBL) algorithm \cite{wipf07,tipping01} when the underlying parameter is multi-dimensional. Contributions in the paper are as follows:\begin{itemize}[leftmargin=*,noitemsep]
    \item A new Bayesian strategy with reduced complexity based on novel \emph{remodeling} and \emph{reparameterization} of the data model is proposed.
    \item An approach that induces a natural coupling of the parameter in the two dimensions by introducing \emph{block sparsity}.
    \item A solution to the two dimensional HR problem using a 1-D grid; learned block structure allows estimating the parameters in other dimension in a \emph{gridless} manner.
    \item Although we solve the problem with a grid in one dimension, we numerically show that our method does not suffer from poor \emph{identifiability}.
\end{itemize}In section~\ref{sec:probdef} we introduce the system model and discuss some background on the problem at hand. 
The proposed method is discussed in section~\ref{sec:propmethod}. Finally in section~\ref{sec:numsim}, we provide numerical results which support the usefulness of the proposed technique.
\section{System Model And Background}\label{sec:probdef}
\subsection{System Model}
Consider a URA with $N_x$ sensors along the $x$ axis and $N_y$ sensors along the $y$ axis. The inter-element spacing used in the array is assumed to be half-wavelength along both the axes, and $L$ measurements or snapshots are collected. The received measurements are composed of $K$ narrowband source signals. We assume that signals impinging on the array come from a finite 2D grid of elevation ($\theta_k\in[0,90]$) and azimuth ($\phi_k\in[0,360)$) angles, where $k\in\{1,\ldots,K\}$, corresponding to the $k$th source. The received measurement at the $(n_x,n_y)$ sensor, at the $l$th instant is given by ($0\leq n_x<N_x,0\leq n_y<N_y;n_x,n_y\in\mathbb{Z}$)\begin{multline}
    \bar{\mathbf{Y}}_l[n_x,n_y]=\sum_{k=1}^Ks_{k,l}\>\mathrm{exp}\left(j\pi\left(n_{x}u_k+n_{y}v_k\right)\right)\\+\bar{\mathbf{V}}_l[n_x,n_y],\label{eq:2dmodlsclr2}
\end{multline}where $u_k=\cos{\phi_k}\sin{\theta_k},v_k=\sin{\phi_k}\sin{\theta_k}$ ($u_k,v_k\in[-1,1],$ $u_k^2+v_k^2\leq 1$). In the above equation, $s_{k,l}$ denotes the transmitted symbol by the $k$th source, and $\bar{\mathbf{V}}_l[n_x,n_y]$ denotes the noise at the $l$th instant modeled as white circular Gaussian. Eq.~(\ref{eq:2dmodlsclr2}) can be re-written in matrix form as:\begin{IEEEeqnarray}{rcl}
    \bar{\mathbf{Y}}_l&=&\sum_{k=1}^Ks_{k,l}\>\mathbf{\Phi}_{u,k}\mathbf{\Phi}_{v,k}^T+\bar{\mathbf{V}}_l\quad (0\leq l< L)\IEEEeqnarraynumspace\label{eq:2dmodlmtx1}\\
    &=&\mathbf{\Phi}_u\mathbf{S}_l\mathbf{\Phi}_v^T+\bar{\mathbf{V}}_l.\IEEEeqnarraynumspace\label{eq:2dmodlmtx2}
\end{IEEEeqnarray}In eq.~(\ref{eq:2dmodlmtx1}), $\mathbf{\Phi}_{u,k}=[1,\mathrm{exp}\left(j\pi u_k\right),\ldots,\mathrm{exp}\left(j\pi(N_x-1)u_k\right)]^T$ and $\mathbf{\Phi}_{v,k}=[1,\mathrm{exp}\left(j\pi v_k\right),\ldots,\mathrm{exp}\left(j\pi (N_y-1)v_k\right)]^T$ represent the array response vectors to the $k$th source along the $x$ and $y$ axes, respectively. Let the grid sizes in the $u$ and $v$-spaces be $M_u$ and $M_v$ respectively. In eq.~(\ref{eq:2dmodlmtx2}) we introduce the corresponding dictionary matrices $\mathbf{\Phi}_u\in\mathbb{C}^{N_x\times M_u}$ and $\mathbf{\Phi}_v\in\mathbb{C}^{N_y\times M_v}$ in the $u$ and $v$ space respectively. The symbol matrix $\mathbf{S}_l$ has at most $K$ non-zero entries corresponding to the $K$ source symbols. In addition to sources located on-grid we also assume the notion of \emph{common-sparsity} along the snapshots, i.e. the source locations $(u_k,v_k)$ although unknown, remain fixed over time. Consequently, the support of the symbol matrix $\mathbf{S}_l$ is constant.

We shall use the above model throughout this paper, as (uniform) grid definitions in $(u,v)-$space often lead to less coherent dictionaries than those in $(\theta,\phi)-$space. The entries in $\mathbf{S}_l$ are assumed to be zero mean and uncorrelated, and the matrices are generated as i.i.d. over the snapshots.
\subsection{Background}
\subsubsection{Using Kronecker Product of Two 1-D Dictionaries}
A straightforward way to recover the support i.e., the $K$ source locations is to extend an off-the-shelf sparse signal recovery algorithm to handle 2D-grid points. For example, we can take the transpose of the received snapshot and vectorize the result to get ($0\leq l<L$)\begin{equation}
    \mathbf{y}_l=\mathrm{vec}(\bar{\mathbf{Y}}_l^T)=(\mathbf{\Phi}_u\otimes\mathbf{\Phi}_v)\mathrm{vec}(\mathbf{S}_l^T)+\mathrm{vec}(\bar{\mathbf{V}}_l^T).
\end{equation}We now concatenate the $L$ snapshots together to get\begin{IEEEeqnarray}{rcl}
\mathbf{Y}&=&(\mathbf{\Phi}_u\otimes\mathbf{\Phi}_v)\left[\begin{array}{cccc}\mathrm{vec}(\mathbf{S}_0^T)&\mathrm{vec}(\mathbf{S}_1^T)&\ldots\>\mathrm{vec}(\mathbf{S}_{L-1}^T)\end{array}\right]\nonumber\\
&& \quad+\left[\begin{array}{cccc}\mathrm{vec}(\bar{\mathbf{V}}_0^T)&\mathrm{vec}(\bar{\mathbf{V}}_1^T)&\ldots&\mathrm{vec}(\bar{\mathbf{V}}_{L-1}^T)\end{array}\right]\nonumber\\
&=&(\mathbf{\Phi}_u\otimes\mathbf{\Phi}_v) \bar{\mathbf{X}}+\mathbf{V}.\label{eq:2DCS}
\end{IEEEeqnarray}Thus, the resulting dictionary to be defined is simply the Kronecker product of the 1-D dictionaries. However such an extension quickly increases the computational cost. The complexity per iteration using the MSBL algorithm \cite{wipf07} is $O(N_x^2N_y^2\times M_uM_v)$ i.e. roughly quadratic in the 1-D grid size. The goal of our work is to reduce this complexity significantly while retaining the superior source identifiability characteristics by preserving the dimension of the measurements. The dimension preserving requirement allows us to identify more sources than what is possible by essentially solving two 1-D problems and then coupling the resulting 1-D solutions.
\subsubsection{On Coupling the Solutions of Two 1-D Problems:}
This method employs one dictionary at a time, and each time considers measurements along one axis and concatenates data along the other axis. For estimating the $u$ components, instead of vectorizing the matrix of measurements obtained at each snapshot, one simply works with the following concatenated matrix
$\left[\bar{\mathbf{Y}}_0,\bar{\mathbf{Y}}_1, \hdots, \bar{\mathbf{Y}}_{L-1}\right].$ When estimating the $v$ components, one utilizes $\bar{\mathbf{Y}}^T_l.$ This method although keeps the computational complexity low, but it suffers from poor identifiability i.e. the number of sources that are recovered. This is because the dimension of the vector measurements is either $N_x$ or $N_y$, much smaller than $N_x N_y$ used in the Kronecker product based approach. It also treats the data along the other dimension as independent snapshots which is not true for the columns within a snapshot (matrix measurement) and leads to sub-optimal results.

\section{Proposed Bayesian Strategy}\label{sec:propmethod}
The goal of this work is to provide a solution with bounded complexity, much lower than using Kronecker product of dictionaries, while exploiting the rich structure in the received measurements. The basic idea in order to lower the computational complexity is to essentially solve for one parameter using a 1-D dictionary,
and at the same time exploiting additional \emph{structure} to simultaneously infer the coupled parameter in the second dimension.
\subsection{Inducing Parameter Coupling via Block Sparsity}
We begin with rewriting eq.~(\ref{eq:2DCS}) and simplifying it further.\begin{IEEEeqnarray}{rcl}
\mathbf{Y}&=&(\mathbf{\Phi}_u\otimes\mathbf{\Phi}_v) \bar{\mathbf{X}}+\mathbf{V}\nonumber\\
&=&\left(\mathbf{\Phi}_u\otimes\mathbf{I}_{N_y}\right)\left(\mathbf{I}_{M_u}\otimes\mathbf{\Phi}_v\right)\Bar{\mathbf{X}}+\mathbf{V}\label{eq:krontrick}\\
&=&\left(\mathbf{\Phi}_u\otimes\mathbf{I}_{N_y}\right)\mathbf{X}+\mathbf{V}\>\>(\mathrm{for\>some\>}\mathbf{X}=\left(\mathbf{I}_{M_u}\otimes\mathbf{\Phi}_v\right)\Bar{\mathbf{X}}),\IEEEeqnarraynumspace\label{eq:vecmodel}
\end{IEEEeqnarray}where in eq.~(\ref{eq:krontrick}) we use the result on Kronecker product: $\mathbf{AC}\otimes\mathbf{BD}=\left(\mathbf{A}\otimes\mathbf{B}\right)\left(\mathbf{C}\otimes\mathbf{D}\right)$. Equation~(\ref{eq:vecmodel}) defines our new model and an important contribution of this work. Our strategy is to treat $\mathbf{D}_u=\mathbf{\Phi}_u\otimes\mathbf{I}_{N_y}$ as the effective dictionary and to first estimate the $u$'s. Note that the row dimension of $\mathbf{D}_u=\mathbf{\Phi}_u\otimes\mathbf{I}_{N_y}$ is the same as in MSBL and this allows the source identifiability properties to be retained. The number of columns is however much smaller leading to the low complexity aspect of the approach.

Another important contribution and unique feature of the method is the estimation of the components in the second dimension. The assumptions made on the symbol matrices $\mathbf{S}_l,0\leq l<L$ translate to the fact that each column of $\bar{\mathbf{X}}$ has uncorrelated components, and the columns themselves are i.i.d.. On the other hand, consider (using MATLAB notations below)\begin{IEEEeqnarray}{rcl}
    \mathbf{X}^i&=&\mathbf{X}((i-1)N_y+1:iN_y,:),\qquad(i\in\{1,\ldots,M_u\})\nonumber\\
    &=&\mathbf{\Phi}_v\bar{\mathbf{X}}((i-1)M_v+1:iM_v,:),\label{eq:blockdef}
\end{IEEEeqnarray}which implies that each such block, $\mathbf{X}^i$, has correlated components along each column. Let $\mathbf{x}^i_l\in\mathbb{C}^{N_y}$ denote the $l$th column of the $i$th block, and let $\mathbf{x}_l=\left[(\mathbf{x}^1_l)^T,\ldots,(\mathbf{x}^{M_u}_l)^T\right]^T$ denote the $l$th column of $\mathbf{X}$. Then eq.~(\ref{eq:blockdef}) implies that $\mathbf{x}^i_l$ has correlated components. Hence, we conclude that $\mathbf{X}$ in the new model i.e. in eq.~(\ref{eq:vecmodel}) is \emph{block sparse}. Also, the correlation within each block is \emph{structured}, and influenced by the coupled parameters in the other dimension i.e. $v$. We dissolve the parameter $v$ from the original model (eq.~(\ref{eq:2DCS})) completely, and instead track the correlation matrices $\mathbf{R_x}_{,i}=\mathrm{E}\left[\mathbf{x}^i_l(\mathbf{x}^i_l)^H\right]$, where $\mathrm{E}[.]$ denotes expectation operation. This \emph{reparameterization} allows us to estimate the coupled $v_i$'s, in a gridless manner. 
Our approach follows a maximum-likelihood (ML) principle, although for the sake of simplicity we do not impose the additional (Toeplitz) correlation structure here and show through numerical simulations that this is adequate. 
\subsection{Model Priors}
As per our observation in eq.~(\ref{eq:blockdef}), we introduce a Gaussian \emph{block} prior on $\mathbf{x}^i_l, i\in\{1,\ldots,M_u\}$ as\begin{equation}
    p(\mathbf{x}^i_l;\gamma_i,\mathbf{B}_i)\sim\mathcal{N}\left(\mathbf{0},\gamma_i\mathbf{B}_i\right),\quad(0\leq l<L) 
\end{equation}such that there is no inter-block correlation. Thus the prior on $\mathbf{x}_l$ is given by\begin{equation*}
    p(\mathbf{x}_l;\gamma_i,\mathbf{B}_i,\forall i)\sim \mathcal{N}\left(\mathbf{0},\mathbf{\Sigma}_0\right),
\end{equation*}\begin{equation}
    \mbox{where, }\mathbf{\Sigma}_0=\left[\begin{array}{ccc}
        \gamma_1\mathbf{B}_1 & & \\
         & \ddots &\\
         & & \gamma_{M_u}\mathbf{B}_{M_u}
    \end{array}\right].
\end{equation} Notice that the prior does not depend on $l$ as the columns are i.i.d over the snapshots. Our formulation here induces a targeted block structure, unlike in T-SBL \cite{zhang11} where a temporal correlation structure exists over all the snapshots. Our approach here is valid for 2-D harmonic retrieval problems and also extends more generally to the MHR problem, as long as the snapshots are independent. The model parameters are estimated using a ML approach and the negative of the log-likelihood function is given by\begin{equation}
    -\log p(\mathbf{Y};\lambda,\gamma_i,\mathbf{B}_i,\forall i)=\log\det\mathbf{\Sigma_y}+\mathrm{tr}\left(\mathbf{\Sigma_y}^{-1}\hat{\mathbf{S}}_{\mathbf{y}}\right),\label{eq:mlcost}
\end{equation}where $\mathbf{\Sigma_y}\triangleq\mathbf{D}_u\mathbf{\Sigma}_0\mathbf{D}_u^H+\lambda\mathbf{I}$, $\lambda$ denotes the noise variance parameter to be estimated, and $\hat{\mathbf{S}}_{\mathbf{y}}=\frac{1}{L}\sum_l\mathbf{y}_l\mathbf{y}_l^H$ denotes the sample covariance matrix. 
\subsection{Algorithm Development}
The cost function in eq.~(\ref{eq:mlcost}) is non-convex and any majorization-minimization framework can be used to minimize the cost function. We here employ the Expectation-Maximization (EM) framework to effectively maximize $p(\mathbf{Y};\lambda,\gamma_i,\mathbf{B}_i,\forall i)$. We begin with estimating the posterior density $p(\mathbf{x}_l\mid \mathbf{y}_l;\lambda^{(k)},\gamma_i^{(k)},\mathbf{B}_i^{(k)},\forall i)$ based on the current estimate of parameters $\mathbf{\Theta}=\left[\lambda,\gamma_i,\mathbf{B}_i,\forall i\right]$. We drop the superscript $(k)$ denoting current iteration for notational simplicity.\begin{equation}
    p(\mathbf{x}_l\mid \mathbf{y}_l;\mathbf{\Theta})\sim\mathcal{N}\left(\bm{\mu}_{\mathbf{x},l},\mathbf{\Sigma_x}\right),
\end{equation}\begin{IEEEeqnarray}{rll}
\mbox{where, }\bm{\mu}_{\mathbf{x},l}&=&\frac{1}{\lambda}\mathbf{\Sigma}_x\mathbf{D}_u^H\mathbf{y}_l\label{eq:meanupdate}\\
\mathbf{\Sigma_x}&=&\mathbf{\Sigma}_0-\mathbf{\Sigma}_0\mathbf{D}_u^H\left(\lambda\mathbf{I}_{N_xN_y}+\mathbf{D}_u\mathbf{\Sigma}_0\mathbf{D}_u^H\right)^{-1}\mathbf{D}_u\mathbf{\Sigma}_0.\label{eq:covupdate}\IEEEeqnarraynumspace
\end{IEEEeqnarray}Note that the mean of the posterior depends on the snapshot index, $l$. We concatenate the posterior mean over snapshots to form $\bm{\mu}_{\mathbf{x}}=\left[\bm{\mu}_{\mathbf{x},0},\ldots,\bm{\mu}_{\mathbf{x},L-1}\right]$.

We shall now proceed to updating the current estimate of the parameters in $\mathbf{\Theta}$. In contrast to the work in \cite{zhang11} where a single correlation matrix $\mathbf{B}$ was rightfully motivated, here we take a different approach. Since the $\mathbf{B}_i$'s are expected to carry information about the coupled $v_i$'s corresponding to individual $u_i$'s, we do not enforce a single $\mathbf{B}$ matrix. This strategy is crucial when identifying sources with distinct $u_i$'s. Following steps are similar to those followed in \cite{zhang11} and we here simply mention the update equations for $\gamma_i$'s, $\mathbf{B}_i$'s and the noise variance estimate $\lambda$.\begin{IEEEeqnarray}{rcl}
\gamma_i&\leftarrow&\frac{1}{N_y}\mathrm{tr}\left(\mathbf{B}_i^{-1}\left(\mathbf{\Sigma_x}^i+\frac{1}{L}\bm{\mu}_{\mathbf{x}}^i\left(\bm{\mu}_{\mathbf{x}}^i\right)^H\right)\right)\label{eq:gammaupdate}\\
\mathbf{B}_i&\leftarrow&\frac{1}{\gamma_i}\left(\mathbf{\Sigma_x}^i+\frac{1}{L}\bm{\mu}_{\mathbf{x}}^i\left(\bm{\mu}_{\mathbf{x}}^i\right)^H\right),\label{eq:Bupdate}
\end{IEEEeqnarray}where $\mathbf{\Sigma_x}^i=\mathbf{\Sigma_x}\left((i-1)N_y+1:iN_y, (i-1)N_y+1:iN_y\right)$ and $\bm{\mu}_{\mathbf{x}}^i=\bm{\mu}_{\mathbf{x}}\left((i-1)N_y+1:iN_y, :\right),i\in\{1,\ldots,M_u\}$. For updating $\lambda$ we use the following update equation:\begin{IEEEeqnarray}{rcl}
\lambda&\leftarrow&\frac{1}{N_xN_yL}\lVert\mathbf{Y}-\mathbf{D}_u\bm{\mu}_{\mathbf{x}}\rVert_F^2\nonumber\\
&&\>+\frac{\lambda}{N_x}\mathrm{tr}\left(\mathbf{\Phi}_u\bm{\Gamma}\mathbf{\Phi}_u^H\left(\mathbf{\Phi}_u\bm{\Gamma}\mathbf{\Phi}_u^H+\lambda\mathbf{I}\right)^{-1}\right)\label{eq:lambdaupdate}\IEEEeqnarraynumspace
\end{IEEEeqnarray}Note that putting $N_y=1$ in the above equations (except for $\lambda$) gives us the update equations for the EM-MSBL algorithm. In the above, we used the notation, $\bm{\Gamma}=\mathrm{diag}\left(\bm{\gamma}\right)=\mathrm{diag}([\gamma_1,\ldots,\gamma_{M_u}])$. We denote the algorithm using the equations~(\ref{eq:meanupdate}),(\ref{eq:covupdate}),(\ref{eq:gammaupdate}),(\ref{eq:Bupdate}) and (\ref{eq:lambdaupdate}) as H-MSBL, emphasizing that our approach is applicable to the Harmonic retrieval problem. A few practical remarks are in order.\vspace{0.1cm}\\{\bf Remark 1:} We normalize the correlation matrices, $\mathbf{B}_i\leftarrow\mathbf{B}_i/\lVert\mathbf{B}_i\rVert_F$ , $\forall i$, so as to separate the role of $\mathbf{B}_i$ and $\gamma_i$. The above update equation for $\lambda$ is different from what is derived using the EM algorithm. The latter was emprirically observed to be unstable. Thus, instead we replaced it with an adaptation of eq. (33) in \cite{zhang11} for high-SNR case and also incorporated suggestions therein for low SNR case which produced stable updates.\\{\bf Remark 2:} Given the number of sources to be identified, the top peaks in the recovered $\bm{\gamma}$ provide information about the $u$ component of the present sources. The corresponding $\mathbf{B}_i$'s provide information about the coupled $v$ component. Thus the need to separately estimate $v$'s and coupling them with $u$'s is prevented. In our simulation, we use root-MUSIC \cite{trees02} to extract the coupled $v$-components.\\{\bf Remark 3:} An interesting challenging scenario in the 2-D HR problem is when the distinct 2-D harmonics have the same value in one dimension. Methods have been designed that assume the harmonics are distinct in both dimensions \cite{rao84} and special care has to be taken in algorithm development to deal with the above mentioned challenging scenario \cite{hua93}. It will be shown numerically that H-MSBL works even in this context when  multiple sources share the same $u$ or $v$ grid point.

\subsection{Computational Complexity: H-MSBL vs MSBL}\label{sec:analysis}
The per iteration complexity of the MSBL algorithm applied to the model in eq.~(\ref{eq:2DCS}) is $O(N_x^2N_y^2\times M_uM_v)$. The same for the proposed H-MSBL algorithm is $O(N_x^2N_y^2\times M_uN_y)$ (or is $O(N_x^2N_y^2\times M_vN_x)$), if the grid is defined on $u$\footnote{this assumes $M_u>N_x$, which is usually the case.} (or on $v$). This is a significant reduction in the computational complexity, as both $N_x$ and $N_y$ are physical array dimensions, and thus are much smaller than the corresponding grid sizes. Like the MSBL algorithm, the complexity of the proposed algorithm does not depend on the number of snapshots. This is because the (ML) cost function depends on the snapshots only through the outer product $\mathbf{YY}^H$. To achieve the proposed complexity, one has to replace $\mathbf{Y}$ with $\tilde{\mathbf{Y}}\in\mathbb{C}^{N_xN_y\times\mathrm{rank}(\mathbf{Y})}$ in the update equations, such that $\mathbf{YY}^H=\tilde{\mathbf{Y}}\tilde{\mathbf{Y}}^H$, and only compute the diagonal blocks of $\mathbf{\Sigma_x}$ of size $N_y\times N_y$.

The MSBL algorithm introduces a missing or latent variable, $\bar{\mathbf{X}}\in\mathbb{C}^{M_uM_v\times N_xN_y}$ which is more \emph{complex} than the latent variable $\mathbf{X}\in\mathbb{C}^{M_uN_y\times N_xN_y}$ we have introduced, both in the reduced dimension. On the other hand, the (ML) cost function in MSBL is defined over $\bm{\gamma}\in\mathbb{C}^{M_uM_v}$, whereas here the cost function is defined over $\bm{\Theta}$ involving a total of $M_u(N_y^2+1)$ parameters, ignoring the noise variance parameter. The EM algorithm is known to converge slowly if
one chooses an overly informative complete data. In the next section we numerically compare the rate of convergence of the two algorithms.

\section{Numerical Simulations}\label{sec:numsim}
In this section, we compare the performance of the proposed H-MSBL algorithm to MSBL (using Kronecker product of 1-D dictionaries in eq.~(\ref{eq:2DCS})). The MSBL is chosen because it is found to be comparable, if not better than most algorithms, when applied to direction of arrival estimation. In addition, it is robust to impairments such as source correlation \cite{pote20}. Furthermore, a competing algorithm like MUSIC also has the same complexity issues as MSBL in the search process. The main purpose of the work is to show that the low complexity H-MSBL can compare favorably with state of the art algorithms and so the more complex MSBL is a good reference point for comparison.

Unless otherwise specified the source location components, $(u_k,v_k),\forall k$, are distinct. The H-MSBL algorithm proceeds by defining grid on the $u$ component, and estimates $v$ in a gridless manner. The number of EM-iterations performed are mentioned alongside the algorithm. For e.g., a MSBL curve using $500$ EM-iterations is denoted by MSBL-$500$. Note that, during the implementation, the combined $(u,v)$-grid size is less than $M_uM_v$ as points that violate the constraint $u^2+v^2\leq 1$ are discarded from the dictionary, $\mathbf{\Phi}$. All algorithms are initialized with $\bm{\gamma}_{\mathrm{init}}=\frac{\lVert\mathbf{YY}^H/L\rVert_F}{\lVert\mathbf{\Phi\Phi}^H\rVert_F}[1,\ldots,1]^T$. All simulation are carried out in MATLAB 9.4.0.813654 in a Windows 10 system using a 2.7 GHz CPU.
\begin{figure}
\begin{tabular}{cc}
\centering
    \hspace{-0.8cm}\includegraphics[width=0.6\linewidth]{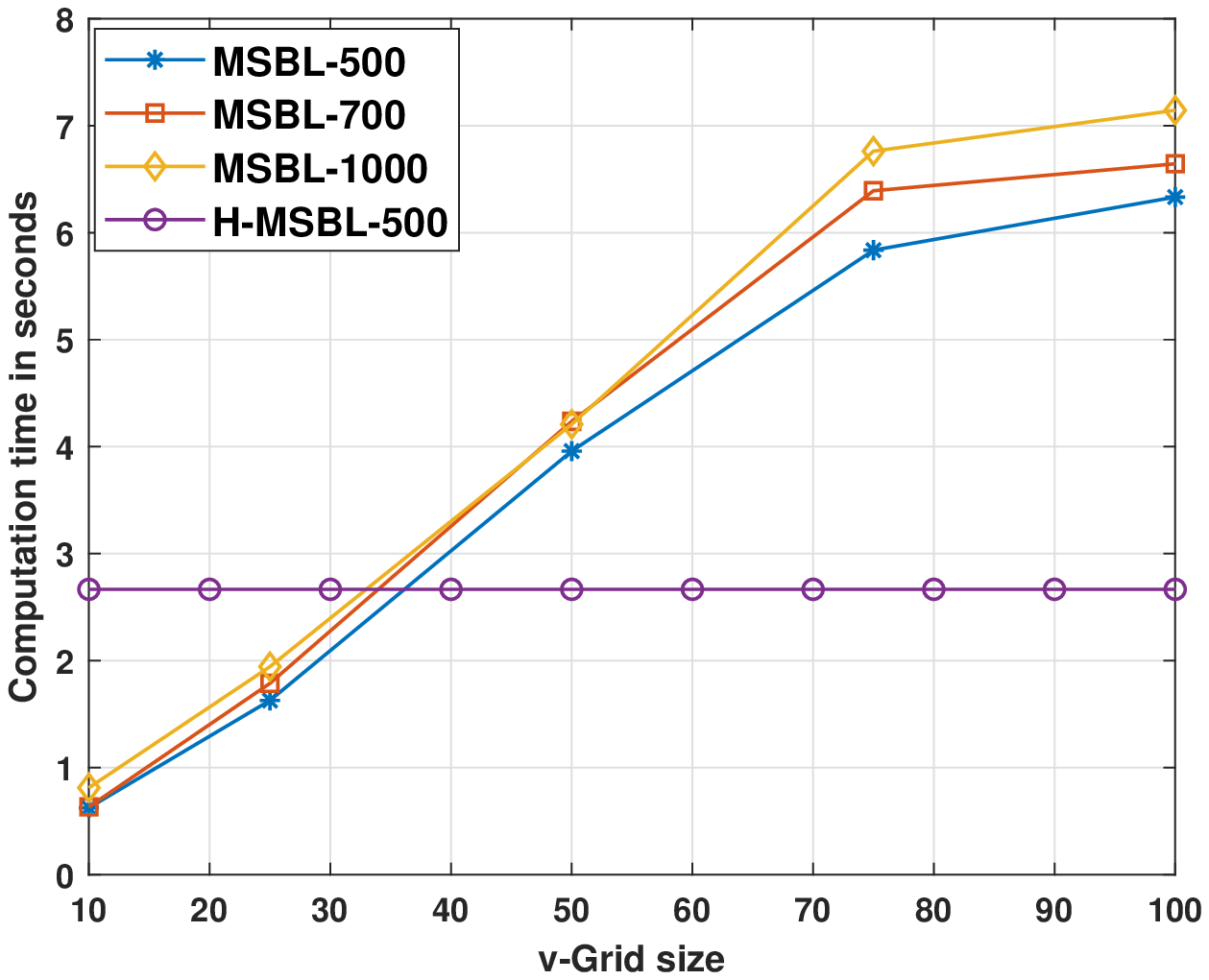} & \hspace{-0.8cm}\includegraphics[width=0.6\linewidth]{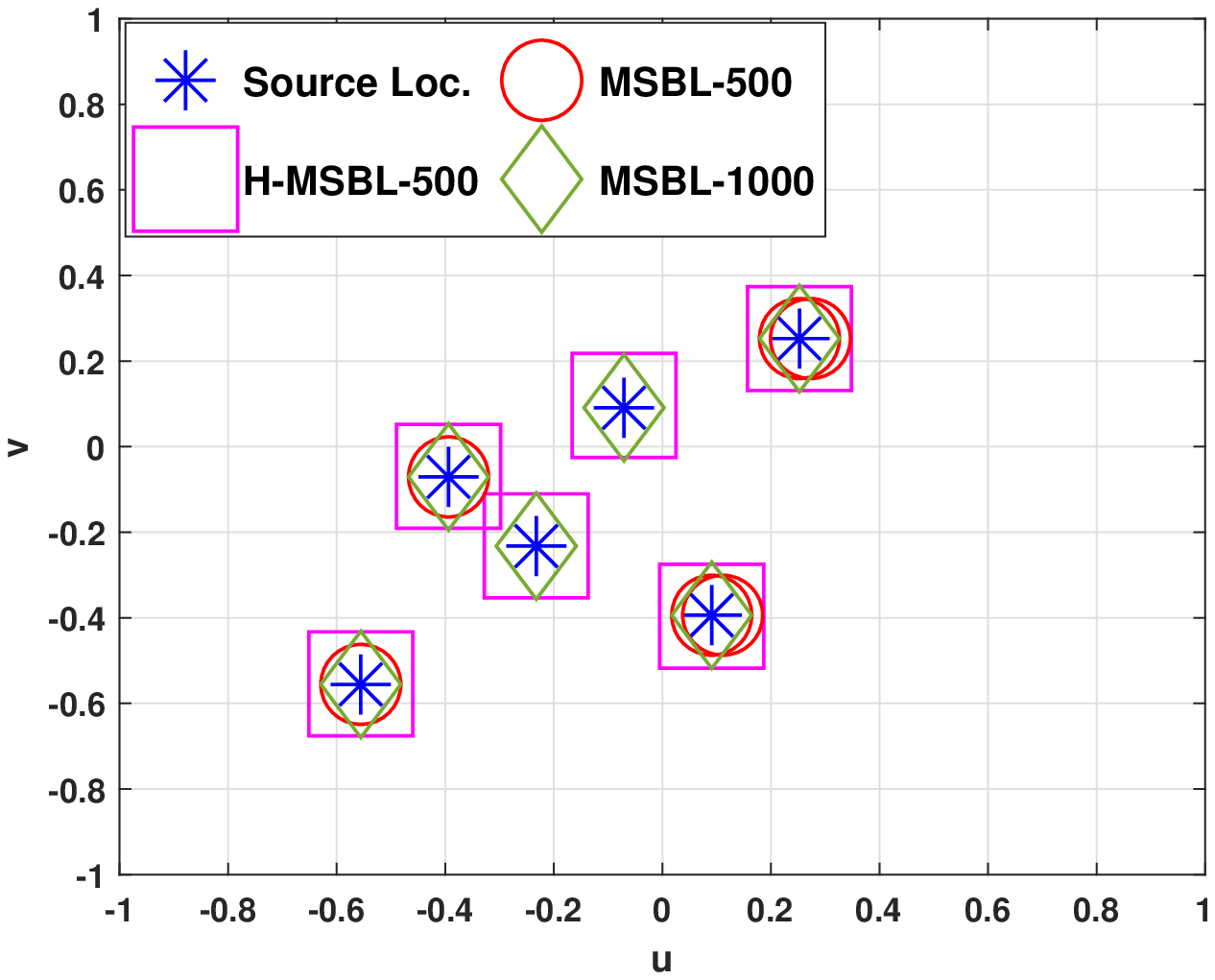} \\
    \hspace{-0.8cm}{\small (a) Computation time vs. $v$-grid size} & \hspace{-0.8cm}{\small (b) Estimated $(u,v)$, $v$-grid size=$100$}
\end{tabular}\caption{H-MSBL vs. MSBL: $N_x=N_y=4,K=6,\mathrm{SNR}=20\mathrm{\>dB},L=50$, $u$-grid size$=100$}\label{fig:complexity}\end{figure}
\begin{figure}\centering \includegraphics[width=0.9\linewidth]{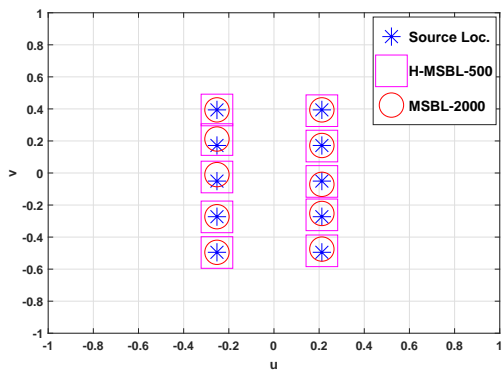}\caption{Comparison on source identifiability: $N_x=3,N_y=6,K=10,\mathrm{SNR}=20\mathrm{\>dB},L=50$, $u,v$-grid sizes$=100$ each}\label{fig:sourceident}\end{figure}

\noindent{\bf Experiment 1:} In this simulation, we wish to compare the computational complexity of using the H-MSBL and MSBL algorithms.   

In fig.~(\ref{fig:complexity}) we compare the computation time in seconds (left) and single instance $(u,v)$ localization performance for H-MSBL and MSBL. As observed in the left plot, the computation time for MSBL grows with the size of the grid on the $v$ component, $M_v$. On the other hand, the same for the H-MSBL algorithm does not depend on $M_v$. For applications such as DoA estimation it is desirable to define a fine grid, but the number of EM-iterations required for convergence also grows. This can be seen in the right plot. Here, H-MSBL is able to localize sources with high accuracy and low complexity, with fewer iterations than MSBL. Note that (non-EM) MSBL implementations may lead to faster convergence, but the complexity will still grow with $M_v$, unlike for H-MSBL.\vspace{0.05cm}\\

\noindent{\bf Experiment 2:} In this simulation we compare the performance of the algorithms when multiple sources share the same $u/v$ grid point.

In fig.~(\ref{fig:sourceident}), we consider a $3\times 6$ array and let sources arrive from a grid of $2$ $u$-points and $5$ $v$-points. This is a challenging scenario as even with $2000$ EM iterations the MSBL estimates are slightly away from true source locations. For H-MSBL, the $v$ component for the multiple sources sharing the same $u$-grid point are estimated using Root-MUSIC on the recovered $\mathbf{B}_i\in\mathbb{C}^{6\times 6}$ matrices. The number of sources to be identified was provided. As can be seen, the proposed technique incurs no loss of source identifiability in this experiment.\vspace{0.05cm}\\
\begin{figure}\centering \includegraphics[width=0.9\linewidth]{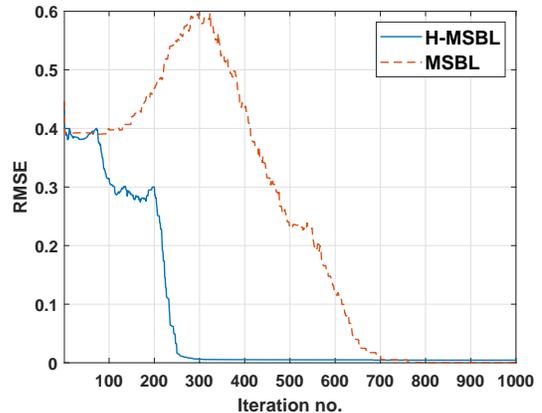}\caption{Rate of convergence: $N_x=4,N_y=4,K=6,\mathrm{SNR}=20\mathrm{\>dB},L=50$, $u,v$-grid sizes$=100$ each}\label{fig:roc}\end{figure}

\noindent{\bf Experiment 3:} As pointed in section~\ref{sec:analysis}, the proposed H-MSBL algorithm introduces a smaller dimensional latent variable $\mathbf{X}$ than MSBL. In this simulation we compare the rate of convergence of the two algorithms.

In fig.~(\ref{fig:roc}) we compare the root mean squared error (RMSE) over the EM iterations for the H-MSBL and MSBL algorithms. The squared error is computed as $\left(u_k-\hat{u}_k\right)^2+\left(v_k-\hat{v}_k\right)^2,\forall k$ and the mean is taken over all sources. As seen from the plot, the proposed algorithm has a faster rate of convergence. Note, the $\gamma_i$'s are pruned if the value falls below $10^{-3}$ for both the algorithms.  
\section{Conclusion}\label{sec:conc}
In this paper, we considered the two dimensional harmonic retrieval (HR) problem and highlighted the increased computational complexity requirements of the MSBL algorithm. We proposed a novel Bayesian strategy by imposing a suitable block sparsity prior. This enables reduction in the effective grid size while maintaining the dimensionality of the measurement space. Thus we are able to identify many more sources than possible by purely 1-D approaches, at the same time achieving reduced complexity. We numerically demonstrated the superior performance of the proposed algorithm.
\bibliographystyle{IEEEbib}
\bibliography{strings,refs}

\end{document}